\documentclass[preprint,12pt]{elsarticle}

\usepackage{xcolor}

\usepackage{amssymb}

\usepackage{hyperref}
    \hypersetup{colorlinks=true,
                linkcolor=red,
                filecolor=blue,
                citecolor=blue,      
                urlcolor=cyan,
    }\usepackage{amsmath}
\usepackage{listings}

\newcounter{bla}

\journal{Computer Physics Communications}

\begin{document}

\begin{frontmatter}



\title{The TDHF code Sky3D version 1.2 }


\author[a]{Abhishek}
\author[a]{Paul Stevenson\corref{author}}
\author[b]{Yue Shi}
\author[a]{Esra Y\"uksel}
\author[c]{A. S. Umar}
\cortext[author] {Corresponding author.\\\textit{E-mail address:} p.stevenson@surrey.ac.uk}

\address[a]{School of Mathematics and Physics, University of Surrey, Guildford, GU2 7XH}
\address[b]{Department of Physics, Harbin Institute of Technology, Harbin 150001, China }
\address[c]{Department of Physics and Astronomy, Vanderbilt University, Nashville, TN 37235, USA}

\begin{abstract}
The Sky3D code has been widely used to describe nuclear ground states, collective vibrational excitations, and heavy-ion collisions. The approach is based on Skyrme forces or related energy density functionals. The static and dynamic equations are solved on a three-dimensional grid, and pairing is been implemented in the BCS approximation. This updated version of the code aims to facilitate the calculation of nuclear strength functions in the regime of linear response theory, while retaining all existing functionality and use cases.   The strength functions are benchmarked against available RPA codes, and the user has the freedom of choice when selecting the nature of external excitation (from monopole to hexadecapole and more). Some utility programs are also provided that calculate the strength function from the time-dependent output of the dynamic calculations of the Sky3D code.

\end{abstract}



\end{frontmatter}



{\bf NEW VERSION PROGRAM SUMMARY}

\begin{small}
\noindent
{\em Program Title: Sky3D}                                          \\
{\em CPC Library link to program files:} (to be added by Technical Editor) \\
{\em Developer's repository link:} \hyperref[]{https://gitlab.surrey.ac.uk/phs3ps/Sky3D/-/tree/v1.2?ref\_type=tags} \\
{\em Code Ocean capsule:} (to be added by Technical Editor)\\
{\em Licensing provisions:} GPLv3  \\
{\em Programming language: Fortran, with one post-processing utility in Python } \\
{\em Supplementary material:}   \\
{\em Journal reference of previous version: Schuetrumpf, B., Reinhard, P. G., Stevenson, P. D., Umar, A. S., and Maruhn, J. A. (2018). The TDHF code Sky3D version 1.1. \href{https://doi.org/10.1016/j.cpc.2018.03.012}{Computer Physics Communications, 229, 211-213.} }                  \\
{\em Does the new version supersede the previous version?: Yes}   \\
{\em Reasons for the new version: The capability of reproducing the nuclear strength function for a variety of newly-coded external boosts has been added. }\\

{\em Nature of the problem:
Calculating nuclear multipole strength functions is a crucial probe that can help model the nuclear system and its structure properties. A variety of models exist for this task, such as QRPA (Quasiparticle Random Phase Approximation) and its variants, but such approaches are often limited due to symmetry constraints. Time-dependent Hartree Fock (TDHF) has been used to simulate nuclear vibrations and collisions between nuclei for low energies without assuming any symmetry in the system. This code extends the TDHF to calculate the multi-pole strength functions of atomic nuclei. We showcase its reliability by comparing it with the established RPA codes for the calculation of such strength functions.}\\

{\em Solution method:
We extended previous versions of the Sky3D code~\cite{MARUHN20142195,SCHUETRUMPF2018211} to include an external boost of multipole type where the user can provide custom input that decides the nature of the multipole (monopole, quadrupole, octupole, and so on) boost. The principal aim is to calculate the multipole strength function, which is the Fourier transform of the time-dependent expectation value of the multipole operator, which has the same form as the external boost. The proper unit conversion is done so that one can extract the exact unit of the thus calculated strength function, which is comparable to the available literature in the field. The boundary conditions are chosen such that a Woods-Saxon-like function cuts off the external field, driving it to zero at the boundary. }\\

{\em Summary of revisions: 
To generate the documentation, go into the Doc-doxygen subdirectory and execute make HTML, make latex, or make all to produce the corresponding version or both of them.
The following is the exhaustive list of changes to the code.
}
\end{small}


\section*{Makefile}
The \textbf{Makefile} has been updated to include targets \textbf{apple} and \textbf{apple\_omp} to use gfortran targetting Apple silicon (ARM-based M1, M2 etc) in which the Lapack dependencies are satisfied with the Apple Accelerate framework and \textbf{fftw} is assumed to be installed via Homebrew\footnote{\texttt{http://brew.sh}}. 

\section*{External Field}
For the dynamic case, the system can be coupled to an external excitation field of general type
\begin{align}
    \label{eq:Ext_operator}
    F_{LM}^{IS}(\vec{r}) &= \sqrt{2L+1} r^L Y_{LM}\\
    F_{LM}^{IV}(\vec{r}) &= \sqrt{2L+1} r^L Y_{LM}f_c,
\end{align}

where IS and IV represent the isoscalar and isovector operators, with some exceptions as noted below.  For the IV case, the factor $f_c$ is multiplied with the operator.  $f_c=\tau_z$ for all the multipole operators except the dipole operator. Here $\tau_z$ is the is the $z$-component of the isospin operator. For the dipole case, it is customary to remove the contribution from the center of mass by setting the $f_c$ to $N/A$ for protons and $-Z/A$ for neutrons.  
For the monopole case, $L = 0$ case, the non-vanishing excitations are produced using 
\begin{align}
    F_{00}^{IS} &= r^2Y_{00} \\
    F_{00}^{IV} &= r^2Y_{00}\tau_z 
\end{align}

We have added this external field in the \textbf{external.f90} file and a new \textbf{ylm.f90} file has been added to calculate the spherical harmonics ($Y_{LM}$) used to define the external field.

To avoid spurious states in the strength function and provide a comparison with the available RPA results, the definition of the external field for the isoscalar dipole excitation is different and defined as~\cite{COLO2013142}

\begin{equation}
\label{eq:dipole}
    F_q(\vec{r}) = \sqrt{3}(r^3 - \frac{5\left\langle r^2 \right\rangle}{3} r)Y_{10}\;,
\end{equation}
where $\left\langle r^2 \right\rangle$ is the expectation value of $r^2$ calculated in the static run.  This includes a higher order ($r^3$ ) term as the natural dipole order induces a shift of the center of mass of the nucleus and not an internal excitation.
The input file contains the variable \textbf{isoext} which decides the nature of the external boost, it should be set to a value of 1 for an isovector type external boost and 0 for the isoscalar type. The code calculates both the isovector and isoscalar multipole response for the either kind of external boost.

\section*{Mix Pairing}
The definition of pairing force is changed slightly to accommodate the implementation of mixed-type pairing by introducing a parameter, $x$, in the case of DDDI input for the ipair=6 value. The new definition, in this case, is 
\begin{equation}
    \label{eq:mix_pair}
    V_P(\vec{r}) = V_0\cdot \xi(\vec{r})\cdot \left( 1 - \frac{x\cdot \rho}{\rho_0} \right).
\end{equation}
The isospin-dependent pairing strength $V_0$ is obtained from the force definition as before and 
\begin{equation}
    \xi(\vec{r}) = \sum_k u_k v_k |\phi_k(\vec{r}) |^2 
\end{equation}
where $\phi_k(\vec{r})$ are the single-particle wave functions.
The value of $x$ is now contained in the \textbf{forces.data} file for the case of DDDI, and by default, the value is given as one to generate the original DDDI case, and a value of 0.5 defines a mix of volume and surface-type pairing. Parameter $x$ is denoted as `\textbf{mixture}' in the code and given as input in the forces.data file along with the \textbf{v0prot} (the strength of pairing for protons in MeV)
\textbf{v0neut} (the strength of pairing for neutrons in MeV) and
\textbf{rho0pr} (the value of $\rho_0$).
Although the same effect of creating a mixed type pairing can be achieved by changing the value of $\rho_0$, e.g., $x = 0.5$, is the same as taking $\rho_0 = 0.32$ without introducing the parameter $x$, but using a parameter like $x$ is easier to read and convey the meaning of a mixture of surface and volume pairing. It is a more common method in the literature~\cite{colò2021user} to use $x$ as a switch, which can go from 0 (volume pairing) to 1 (surface pairing) and fixing the $\rho_0$ to 0.16 fm$^{-3}$.

\section*{Input Files}
The input file for the dynamic run to calculate the time-dependent multipole moments now requires the $L$ and $M$ values used in the definition of the external field [Eq.~(\ref{eq:Ext_operator})]. The main changes are made to the namelist \textit{extrn} as it now includes 

\begin{verbatim}
NAMELIST/extern/ ampl_ext,L_val,M_val,radext,widext,isoext,&
    ipulse,omega,tau0,taut,textfield_periodic,r_avg,only_P
\end{verbatim}
Where \textbf{ampl\_ext} is the strength of external boost, \textbf{L\_val} and \textbf{M\_val} are the $L$ and $M$ values and $r\_avg$ is the value of $\sqrt{\left\langle r^2 \right\rangle}$ also known as the root mean square value (rms) and is used in the isoscalar dipole case [Eq.~(\ref{eq:dipole})]. \\
The variable \textbf{only\_P} is introduced in order to calculate the electromagnetic response, which is obtained by only boosting the proton wavefunctions with the external field (if \textbf{only\_P} = 1). The default value is set to zero. This can be helpful in the evaluation of $B(E\lambda)$ values. However, caution is needed as it can shift the system's center of mass, leading to unreliable results. 

\section*{Output Files}
For every dynamic run, the code now prints the expectation values of all the multipole operators as a function of time in the \texttt{ .res} files. For example, the monopole moment will be printed in \texttt{monopoles.res}, dipole moment in \texttt{dipoles.res}, and so on. 

The corresponding multipole moment file should be considered to calculate the strength function for the given multipole boost.  By analysing a different multipole response to that applied as a boost, one can also explore mode coupling, which appears naturally in a TDHF framework \cite{shi_isoscalar_2023}.

\section*{Utilities}
A new Python program, Fourier.py, is added in the utility section in the `\textbf{utils/strength\_calculation}' directory. The numpy package is required to run this utility program. It calculates the strength function by performing the Fourier transform ($\hat{O}(\omega)$) of the time-dependent multipole moment generated after a boost with a strength $\eta$. The strength function is defined as

\begin{equation}
    \label{eq:strength_function}
    S(E) = \frac{1}{\eta\hbar c\pi} \text{Im}[\hat{O}(\omega)]
\end{equation}

A finite time simulation introduces some artifacts in the Fourier transform if the time signal doesn't die at the end of the simulation~\cite{press2007numerical}. This can be overcome by manually damping the signal by introducing a filter. The filtering can be done in many ways and we provide two filters in this utility program. 
One is named the Cos filter, and the other is the Exponential (Exp) filter, whose definitions are
\begin{align}
    \text{Cos Filter} &= \mathrm{cos}^{n}\left(\frac{\pi t}{2T_0}\right),\\
    \text{Exp Filter} &= \exp{\left(\frac{-\Gamma_0 t}{2\hbar}\right)},\label{eq:exp_fil}
\end{align}

where $\Gamma_0$ and $n$ are parameters whose default values are chosen as 1 MeV and 6, respectively. The Cos filter was initially provided in the previous version of the code.  In the present work, the exponential filter is added to help compare the results with the other codes.  The output of this strength calculation program is saved in two files, one of which contains the isoscalar strength function and other contains the isovector strength function.  The multipole is selected when running the code.

\section*{Benchmarks}
This section compares the isoscalar (IS) and isovector (IV) strength functions obtained for different multipole operators with the RPA results obtained using the hfbcs-qrpa (v1) code\cite{colò2021user}. We compare the results for $^{208}$Pb. In all cases, TDHF satisfies the energy-weighted sum rules (EWSR) with a percentage score of 100$\pm$3 \%.

The M is set to zero in Eq.~(\ref{eq:Ext_operator}) for all multipole operators in this comparison.
A sample of the input file for the dynamic run looks like
\begin{verbatim}
 &files wffile=`restart'/
 &force name=`Sly5', pairing=`NONE' /
 &main mprint=10,mplot=0,
  mrest=500000,writeselect=`rc',
  imode=2,tfft=T,nof=1 /
 &grid nx=24,ny=24,nz=24,dx=1,dy=1,dz=1,
	periodic=F /
 &dynamic nt=50000, dt=0.2, mxpact=4, texternal=T/
 &extern ipulse=0,isoext=0,ampl_ext=5D-5,L_val=0,M_val=0,
   radext=200D0, widext=1D0,textfield_periodic=F /
 &fragments filename=1`./Pb208',fix_boost=T,
      fcent(1,1)=0,0,0    /
\end{verbatim}

\begin{figure}
    \centering
    \includegraphics[width=\textwidth]{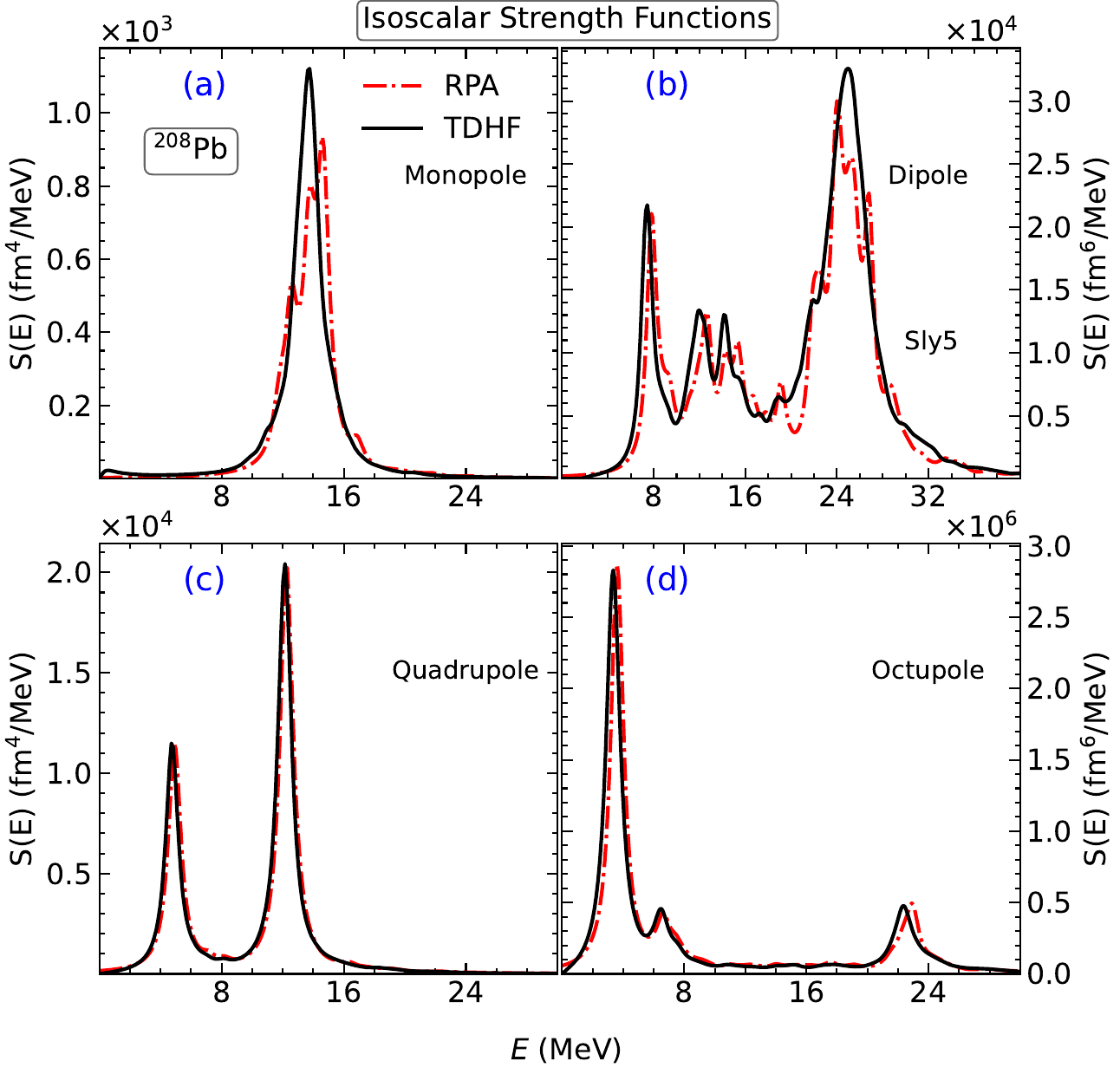}
    \caption{Comparison between the IS strength obtained with SLy5 force using TDHF (Solid Black) and RPA  strength functions (Red dashed) associated with the isoscalar operators, in the case of $^{208}$Pb .}
    \label{fig:pb_poles_IS}
\end{figure}

\begin{figure}
    \centering
    \includegraphics[width=\textwidth]{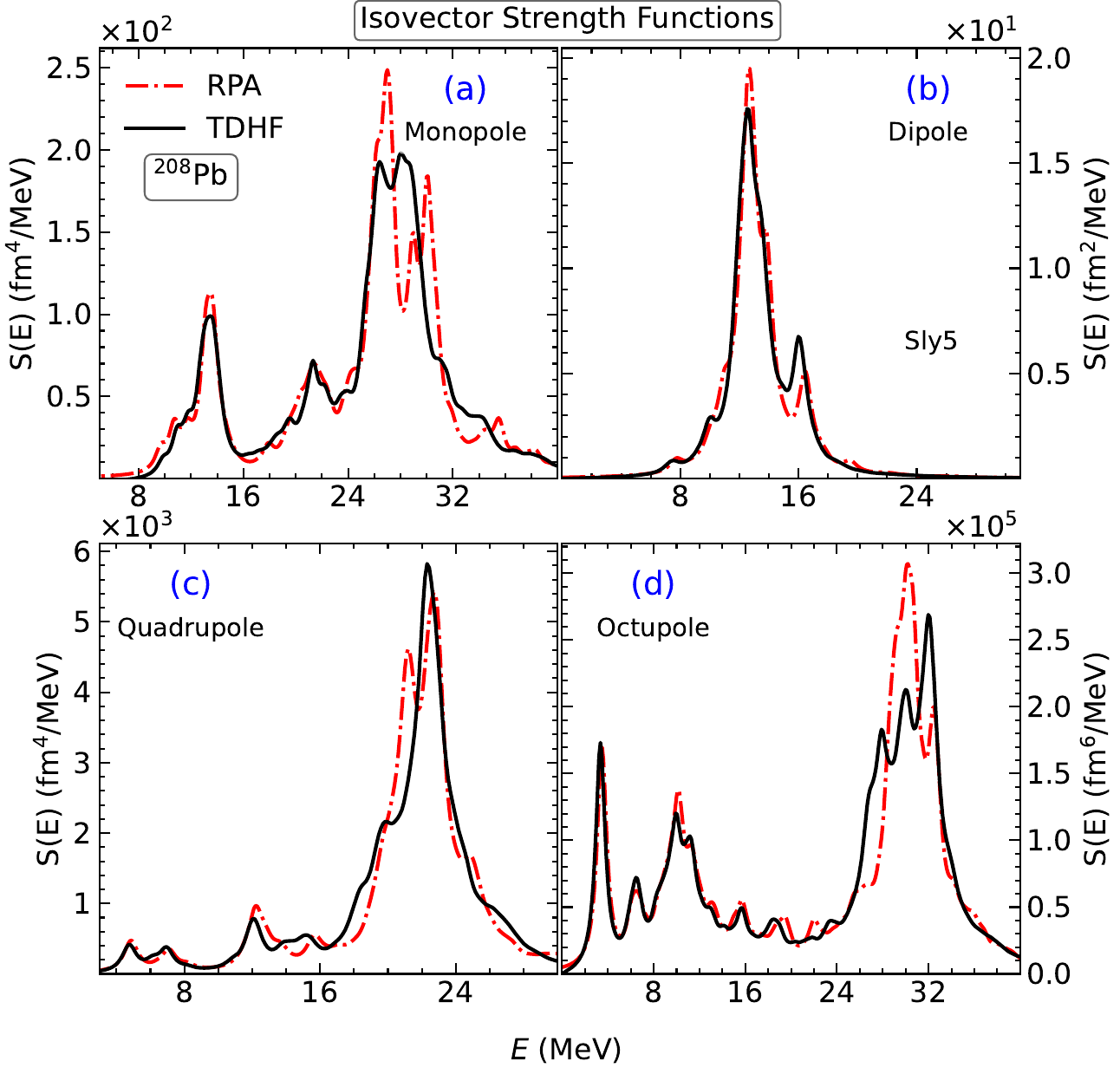}
    \caption{Same as the Fig.~\ref{fig:pb_poles_IS} but with Isovector strength functions.}
    \label{fig:pb_poles_IV}
\end{figure}

Fig.~\ref{fig:pb_poles_IS}, shows the TDHF strength functions for isoscalar monopole to isoscalar octupole boosts obtained with the present version of Sky3D compared with the RPA strength functions in the case of $^{208}$Pb. For the isovector cases, the results are shown in Fig.~\ref{fig:pb_poles_IV}. All the results are obtained using the SLy5 force and the Exp. filter [Eq.~\ref{eq:exp_fil}] with $\Gamma_0 = 1$ MeV is used. Overall, the TDHF results show a good agreement with the QRPA results in all the multipole responses in both isoscalar and isovector cases.  Some differences in which the discretisation of broad peaks above particle decay threshold can be seen.  This is presumably due to boundary conditions at large $r$, which differ between TDHF (wave functions go to zero at the edge of coordinate space box) and RPA (truncated oscillator basis).

\section*{Acknowledgements}
This work has been supported by the UK Science and Technology Facilities Council (STFC) under grants ST/V001108/1.  This work was in part performed using the DiRAC Data Intensive service at Leicester, operated by the University of Leicester IT Services, which forms part of the STFC DiRAC HPC Facility (www.dirac.ac.uk). The equipment was funded by BEIS capital funding via STFC capital grants ST/K000373/1 and ST/R002363/1 and STFC DiRAC Operations grant ST/R001014/1. DiRAC is part of the National e-Infrastructure.  This work is also Sspported by the National Natural Science Foundation of China (12075068, 11705038), the Fundamental Research Funds for the Central Universities (HIT.BRET.2021003).  This work is also supported by the US DOE under grant No. DE-SC0013847

 \clearpage

\bibliographystyle{elsarticle-num}
\bibliography{bib}







\end{document}